\documentclass[english,manuscript,superscriptaddress]{revtex4}
\usepackage[T1]{fontenc}
\usepackage[latin9]{inputenc}
\usepackage{color}
\usepackage{graphicx}

\makeatletter
\@ifundefined{textcolor}{}
{%
 \definecolor{BLACK}{gray}{0}
 \definecolor{WHITE}{gray}{1}
 \definecolor{RED}{rgb}{1,0,0}
 \definecolor{GREEN}{rgb}{0,1,0}
 \definecolor{BLUE}{rgb}{0,0,1}
 \definecolor{CYAN}{cmyk}{1,0,0,0}
 \definecolor{MAGENTA}{cmyk}{0,1,0,0}
 \definecolor{YELLOW}{cmyk}{0,0,1,0}
 }

\makeatother

\usepackage{babel}
\begin{document}

\title{\textcolor{black}{An Approximation to the Cross Sections of $Z_{l}$
Boson Production at CLIC by Using Neural Networks}}

\author{\textcolor{black}{S. Akkoyun}}

\email{sakkoyun@cumhuriyet.edu.tr}

\selectlanguage{english}%

\affiliation{\textcolor{black}{Cumhuriyet University, Faculty of Science, Physics
Department, Sivas, Turkey}}

\author{\textcolor{black}{S.O. Kara}}

\email{sokara@science.ankara.edu.tr}

\selectlanguage{english}%

\affiliation{\textcolor{black}{Ankara University, Physics Department, Ankara,
Turkey}}
\begin{abstract}
\textcolor{black}{In this work, compatible with our previous study,
mass ($M_{Z_{l}}$) and interaction constant ($g_{l}$) of massive
leptonic (leptophilic) boson ($Z_{l}$) at CLIC were investigated
by using artificial neural networks (ANNs). Furthermore, it was seen
that invariant mass distributions for final muons at CLIC after e$^{+}$e$^{-}$$\rightarrow\gamma,Z,Z_{l}\rightarrow\mu^{+}\mu^{-}$
signal e$^{+}$e$^{-}$$\rightarrow\gamma,Z\rightarrow\mu^{+}\mu^{-}$
background processes were consistently predicted by using ANN. Lastly,
for these highly nonlinear data, we have constructed consistent empirical
physical formulas (EPFs) by appropriate feed-forward ANN. These ANN-EPFs
can be used to derive further physical functions which could be relevant
to studying for leptophilic $Z_{l}$ vector boson.}
\end{abstract}
\maketitle

\section{Introduct\i{}on}

The gauging of the baryon and lepton numbers has a long history. In
1955 Lee and Yang proposed massless baryonic \textquotedblleft{}photon\textquotedblright{}
\cite{Lee}, later in 1969 Okun considered massless leptonic \textquotedblleft{}photon\textquotedblright{}
\cite{Okun} in analogy with the baryonic photon. On the other hand
gauging of $B-L$ \cite{Buchmuller,Khalil} is natural in the framework
of Grand Unification Theories. Manifestations of the $Z^{'}$ boson
of the minimal $B-L$ model at future linear colliders and LHC have
been considered in recent paper \cite{Basso}. In \cite{Kara} we
have considered phenomenology of massive U(1) boson coupled to lepton
charge. In this paper, by using data from our previous paper we have
obtained some limit values for \textcolor{black}{$M_{Z_{l}}$} and
$g_{l}$ via artificial neural networks (ANNs).

\textcolor{black}{The physical phenomena involved in massive leptonic
(leptophilic) boson ($Z_{l}$) are characteristically highly nonlinear.
Therefore, it may be difficult to construct explicit form of empirical
physical formulas (EPFs) relevant to $Z_{l}$. Then, by various appropriate
operations of mathematical analysis, derivation of potentially useful
highly nonlinear physical functions for $Z_{l}$ is of utmost interest.
Compatibly a previous theoretical treatment \cite{Y=000131ld=000131z},
appropriate EPFs relevant to $Z_{l}$ can be built by using a feed-forward
artificial neural network (ANN). As we give more details in Section
II , the ANN is a universal nonlinear function approximator \cite{Hornik}.}

Recently, ANNs have emerged with successful applications in many fields,
including Higgs boson search \cite{Acciari,Boos,Hakl,Mellado,Mjahed}.
In this study, compatible with our previous study \cite{Kara}, mass
(\textcolor{black}{$M_{Z_{l}}$}) and interaction constant ($g_{l}$)
of $Z_{l}$ at CLIC were investigated by using ANNs. Besides, the
invariant mass distributions for final muons ($M_{\mu^{+}\mu^{-}}$)
at CLIC with $\sqrt{s}$$=3TeV$ for signal and background were consistently
obtained by using ANN. In all calculations we have performed, signal
and background processes were \textcolor{black}{e$^{+}$e$^{-}$$\rightarrow\gamma,Z,Z_{l}\rightarrow\mu^{+}\mu^{-}$}
and \textcolor{black}{$e^{+}e^{-}\rightarrow\gamma,Z\rightarrow\mu^{+}\mu^{-}$}
respectively. Also, we particularly aim to construct explicit mathematical
functional form of ANN-EPFs for nonlinear data relevant to $Z_{l}$.
While the calculated data were intrinsically highly nonlinear, even
so train set ANN-EPFs successfully fitted these data. Furthermore,
test set ANN-EPFs consistently predicted the data. That is, the physical
laws embedded in the data were extracted by the ANN-EPFs.

\section{ANN and ANN-EPF}

The fundamental task of the artificial neural networks (ANNs) is to
give outputs in consequence of the computation of the inputs. ANNs
are mathematical models that mimic the human brain. They consist of
several processing units called neurons which have adaptive synaptic
weights \cite{Haykin}.ANNs are also efficient tools for pattern recognition.
The ANN consists of three layers named as input, hidden and output
(Fig.1). The number of hidden layers can differ, but a single hidden
layer is enough for efficient nonlinear function approximation \cite{Hornik}.
In this study, one input layer with one neuron, one hidden layer with
many ($h$) neuron and one output layer with one neuron ($1-h-1$)
ANN topology was used for investigation of massive leptonic (leptophilic)
boson ($Z_{l}$) at CLIC with $\sqrt{s}$$=3TeV$ . Analyses were
performed for most convenient hidden neuron numbers in each. The total
numbers of adjustable weights are calculated by using formula given
in ($1$),

\textcolor{black}{
\begin{equation}
(p\times h+h\times r=h\times(p+r)=2h)
\end{equation}
}where p and r are the input and output neuron numbers, respectively.

\begin{figure}

\includegraphics[bb=0bp 0bp 300bp 250bp,scale=0.5]{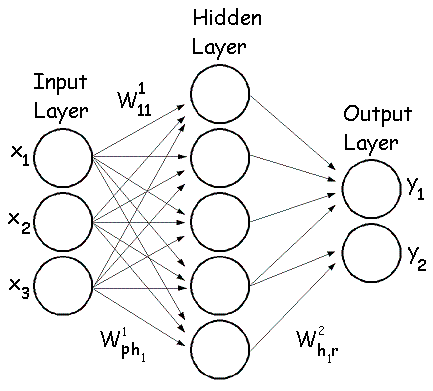}\caption{Fully connected one input-one hidden-one output layer ANN. $x$$_{i}$
and $y$$_{i}$ are input and output vector components, respectively.
Circles are neurons and arrows indicate adaptable synaptic weights.
$w_{jk}^{i}$ : weight vector component, where $i$ is a layer index,
$jk$ weight component from the $jth$ neuron of $ith$ layer and
to $kth$ neuron of ($i+1$)$th$ layer.}

\end{figure}

The neuron in the input layer collects the data from outside and transmits
via weighted connections to the neurons of hidden layer which is needed
to approximate any nonlinear function. The hidden neuron activation
function can be any well-behaved nonlinear function. In this study,
the type of activation functions for hidden layers were chosen as
hyperbolic tangent ($2$). Note that instead of this function, any
other suitable sigmoidal function could also be used. Finally, the
output layer neurons return the signal after the analysis.

\textcolor{black}{
\begin{equation}
tanh=\frac{e^{x}-e^{-x}}{e^{x}+e^{-x}}
\end{equation}
}

An ANN software NeuroSolutions v6.02 was used for separate applications.
In these applications, ANN inputs were $M_{Z_{l}}$ , $g_{l}$ and
$M_{\mu^{+}\mu^{-}}$ and corresponding desired outputs were cross
sections for each input. For all ANN processing case, the data were
divided into two equal separate sets. One of these (\textcolor{black}{$50\%$})
belongs to the training stage and the rest ($50\%$) belongs to the
test stage. In the training stage, a back-propagation algorithm with
Levenberg- Marquardt for the training of the ANN was used. The maximum
epoch number (one complete presentation of the all input-output data
to the network being trained) was $1000$. ANN modifies its weights
by appropriate modifications until an acceptable error level between
predicted and desired outputs is reached. The error function which
measures the difference between outputs was mean square error (MSE)
as given in ($3$),

\textcolor{black}{
\begin{equation}
MSE=\frac{[{\displaystyle {\textstyle \sum_{k=1}^{r}\sum_{i=1}^{N}}(y_{ki}-f_{ki})^{2}]}}{N}
\end{equation}
}where N is the number of training or test samples, $y_{ki}$ and
$f_{ki}$ are the desired output and network output, respectively.
Then by using ANN with final weights, the performance of the network
is tested over test data which are never seen before by network. If
the predictions of the test data are well enough, the ANN is considered
to have consistently learned the functional relationship between input
and output. In this work, the MSE values were between $8$$\times10^{-38}$
and $5\times10^{-3}$ for the training stage and $6\times10^{-11}$
and $4\times10^{-2}$for the test stage. In Fig.2, the training MSE
values for invariant mass distribution of final muons at CLIC with
$M_{Z_{l}}=1TeV$ were given as an illustration.

\begin{figure}

\includegraphics[scale=0.5]{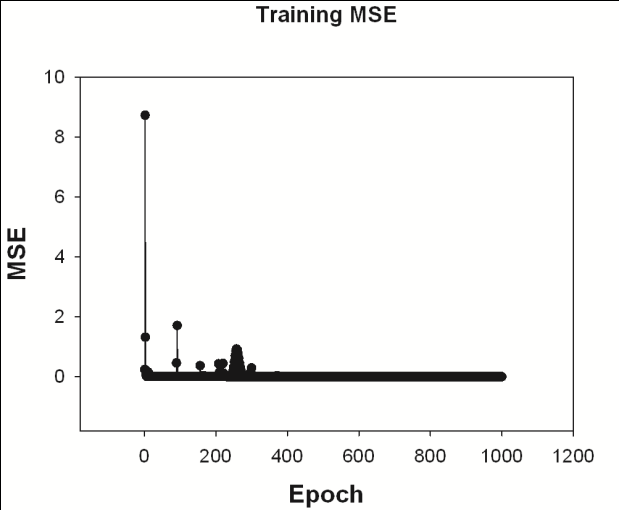}\caption{For invariant mass distribution of final muons at CLIC, the training
MSE values versus epoch number}
\end{figure}

Owing to the fact that a single hidden layer feed-forward ANN is enough
for nonlinear function approximation \cite{Hornik}, in this paper
we used single hidden layer ANNs as previously stated. Here, we only
explain the single hidden layer feed-forward ANN functionality. Borrowing
from \cite{Hornik}, the desired output vector $\vec{y}$ is approximated
by a network multi-output vector $\vec{f}$ which is defined by ($4$).

\textcolor{black}{
\begin{equation}
\vec{f}:R^{p}\rightarrow R^{r}:\vec{f_{k}}(\vec{x})=\sum_{j=1}^{N}\beta_{j}G(A_{j}(\vec{x}));\vec{x}\mathcal{\in}R^{p},\beta_{j}\mathcal{\in}R,A_{j}\mathcal{\in}A^{p},k=1,...,r
\end{equation}
}where $\vec{x}$ is the ANN input vector, $A^{p}$ is the set of
all functions of $R^{p}\rightarrow R$ defined by $A(\vec{x})=\vec{w}\cdot\vec{x}+b$,
$\vec{w}$ is input to hidden layer weight vector, $b$ is the bias
weight. In Fig. 1, the columns of the weight matrices $w^{1}$ and
$w^{2}$ correspond to weight vectors defined in $A(\vec{x})$ and
$\vec{\beta}$ in ($4$). However, as can be seen in Fig. 1 and ($4$),
the correspondences $w^{1}\rightarrow A(\vec{x})$ and $w^{2}\rightarrow\vec{\beta}$
are valid only for single hidden layer feed-forward ANN.

Since a deterministic or random EPF is usually a mathematical vector
function $\vec{y}:R^{p}\rightarrow R^{r}$ between the physical variables
under investigation, particularly ANN is relevant to EPF construction.
Therefore, being a general input-output function estimator, the ANN
defined by ($4$) is particularly relevant in this context. But, although
there can be several independent variables ($p>1$ in ($4$)), the
number of the dependent variables is usually one. Train data for both
independent and dependent physical variables are presented to the
input and output layers respectively. Then after an appropriate weight
adaptation process, the LFNN estimates the unknowable generally nonlinear
EPF.

\section{The concrete algorithm for ANN-EPF construction}

To construct appropriate EPF for highly nonlinear cross sections,
we used one neuron output ANN vector function $\vec{f}$ in ($4$).
However, due to the fact that it gives only the rough structure of
the ANN without generating the final EPF parameters/final ANN optimal
weights, this equation is not sufficient for the complete construction
of the desired nonlinear EPF. In order to obtain the final weight
vector $\vec{w}_{f}$ and the corresponding ANN output vector function
$\vec{f}_{min}=\vec{f}(\vec{w}_{f})$ of ($4$), we simultaneously
used the ($3$) and ($4$). More clearly, given the desired input-output
experimental data, $\vec{f}_{min}$ is the network output vector function
giving the minimum MSE by a convenient ANN weight adaptation. Note
that, $\vec{f}_{min}$ is the best nonlinear estimation vector of
the theoretically unknown desired output function $\vec{y}:R^{p}\rightarrow R^{r}$
(see Fig. 1). In other saying, the unknown vector function $\vec{y}$
is estimated by $\vec{f}_{min}$ which is actually desired nonlinear
EPF that we aim to eventually obtain. $\vec{f}_{min}$ totally depends
on the structure of the network output vector function $\vec{f}$
and the final weight vector $\vec{w}_{f}$ . In ($4$), components
of the weight are embedded in $A(\vec{x})$ and $\vec{\beta}$ . In
($4$), $\vec{f}$ depends on the apparent forms of $A$ and hidden
layer activation. In this paper, setting $\vec{\beta}=w^{2}$ of Fig.
1, hidden layer activation function is nonlinear tangent hyperbolic
and $A$ is the dot product of $w^{1}$and $\vec{x}$ of Fig. 1. So,
we can construct explicit form of $\vec{f}$ . Afterwards, by minimization
of ($4$), we finally obtain $\vec{f}_{min}=\vec{f}(\vec{w}_{f})$
. Now, the concrete ANN-EPF construction algorithm for nonlinear cross
sections is completed. The actual ANN-EPFs results are given in Section
IV.

\section{Results and discussion}

\subsection{ANN-EPFs for train set fittings}

During all the training stages, the number of data points was $50\%$
of all data. For a single hidden layer ANN, the train set nno (neural
network output) fittings for cross sections versus $M_{Z_{l}}$ for
different $g_{l}$ values ($0.10,0.20,$ and $0.30$) were given in
Fig.3. Here the best fitting was obtained for $h=3$ (h: hidden layer
neuron number). In Fig.4 and 5, the train set nno fittings for cross
sections versus $g_{l}$ and $M_{\mu^{+}\mu^{-}}$ were given, respectively.
For $M_{\mu^{+}\mu^{-}}$, two different $M_{Z_{l}}$ values were
used and hidden layer neuron number which gives the best fitting is
$7$. In order to show effect of varying $h$, not only best one but
also different ones of $h$ were given for$g_{l}$ in Fig.5. It can
be clearly seen in the figures, the nno fittings agree exceptionally
well with highly nonlinear calculated data. Additionally, it is clear
in Fig.5 that signal is well above the background.

\begin{figure}

\includegraphics[scale=0.5]{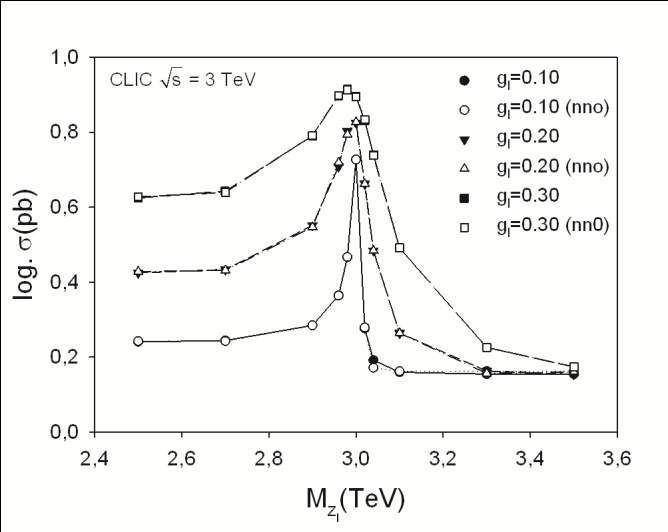}\caption{Calculated and nno train set fittings of cross section versus $M_{Z_{l}}$
for different $g_{l}$ values at CLIC with $\sqrt{s}$$=3TeV$}

\end{figure}

\begin{figure}

\includegraphics[scale=0.5]{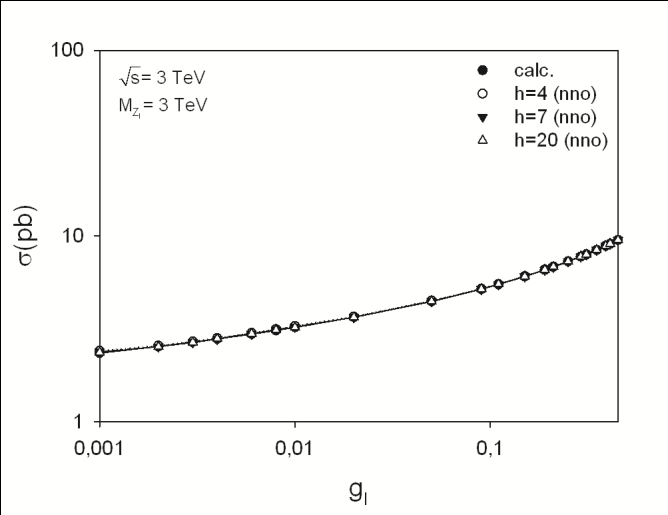}\caption{Calculated and nno train set fitting of cross section versus $g_{l}$
at CLIC with $\sqrt{s}$$=3TeV$ for different hidden layer neuron
number.}
\end{figure}

\begin{figure}

\includegraphics[scale=0.5]{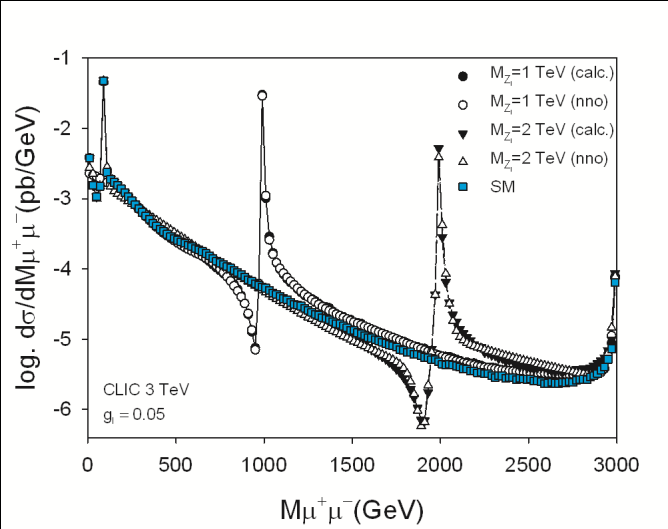}\caption{Calculated and nno train set fittings of differential cross section
versus $M_{\mu^{+}\mu^{-}}$ for SM background and signal at CLIC
with $\sqrt{s}$$=3TeV$.}

\end{figure}

\subsection{Consistency of the constructed ANN-EPFs: Test set predictions}

Unless the train set ANN-EPFs are tested over cross section data,
these fitted EPFs cannot be used consistently over a desired range
of cross section values. If the predictions are consistent with the
test data values, then the ANNs can be taken as appropriate ANN-EPFs.
The corresponding test set nno predictions of Figs.3-5 were given
in Figs.6-8. The number of data points was $50\%$ of all data. As
can be seen in Figs.6-8, the nno predictions agree exceptionally well
with highly nonlinear experimental values. This obviously indicate
that the test set ANNs of cross sections versus $M_{Z_{l}}$ , $g_{l}$
and $M_{\mu^{+}\mu^{-}}$ have consistently generalized the train
ANN fittings. So that, obtained ANNs can safely be used as ANN-EPFs
since the physical law embedded in cross sections versus $M_{Z_{l}}$
, $g_{l}$ and $M_{\mu^{+}\mu^{-}}$ data has been successfully extracted
by the constructed ANN.

\begin{figure}

\includegraphics[scale=0.5]{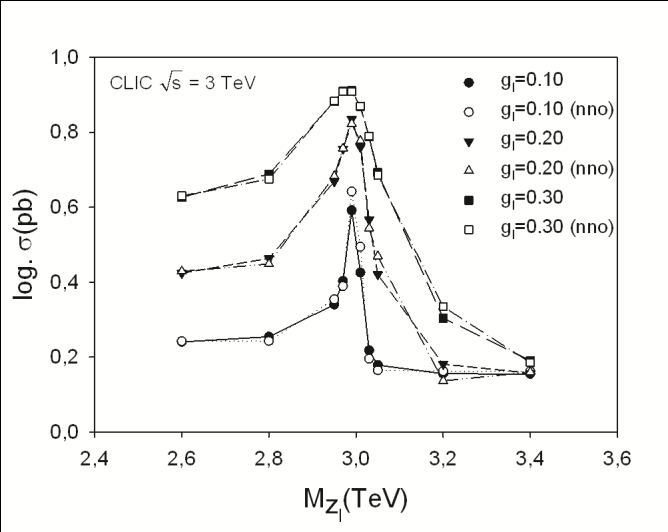}\caption{Calculated and nno test set predictions of cross section versus $M_{Z_{l}}$
for different $g_{l}$ values at CLIC with $\sqrt{s}$$=3TeV$}
\end{figure}

\begin{figure}

\includegraphics[scale=0.5]{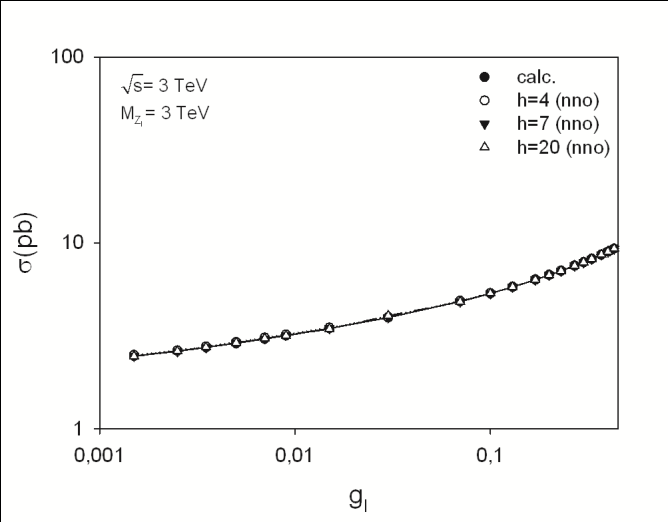}\caption{Calculated and nno test set prediction of cross section versus $g_{l}$
at CLIC with $\sqrt{s}$$=3TeV$ for different hidden layer neuron
number.}

\end{figure}

\begin{figure}
\includegraphics[scale=0.5]{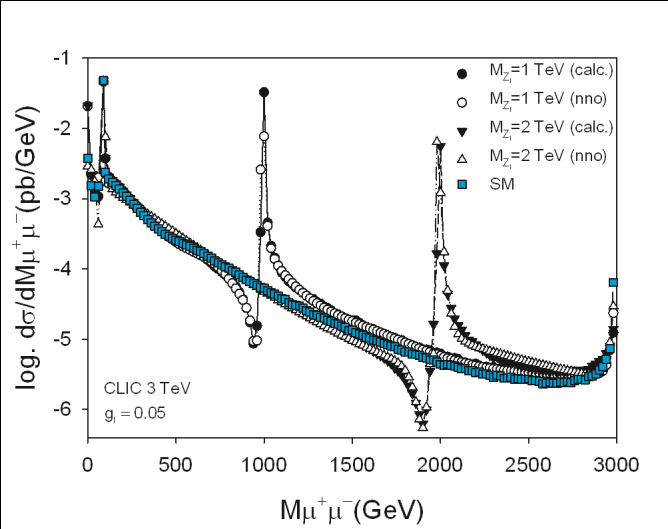}\caption{Calculated and nno test set predictions of differential cross section
versus$M_{\mu^{+}\mu^{-}}$ for SM background and signal at CLIC with
$\sqrt{s}$$=3TeV$ .}

\end{figure}

\section{Conclus\i{}on}

Future linear colliders, like CLIC, will give a chance for investigation
leptophilic vector boson with masses up to the center of mass energy
if $g_{l}\geq10^{-3}$ . It was clearly seen that, ANN method is consistent
with simulations. For highly nonlinear cross sections for $M_{Z_{l}}$
, $g_{l}$ and $M_{\mu^{+}\mu^{-}}$, we have constructed consistent
empirical physical formula (EPFs) by appropriate ANNs. The test set
ANNs of cross sections versus $M_{Z_{l}}$ , $g_{l}$ and $M_{\mu^{+}\mu^{-}}$
have generalized the train ANN fittings. Therefore, the test set ANNs
can be surely used as ANN-EPFs since the physical laws embedded in
cross sections versus $M_{Z_{l}}$ , $g_{l}$ and $M_{\mu^{+}\mu^{-}}$
data have been successfully extracted by the ANN.

\end{document}